\documentclass{aa}
\usepackage{graphicx}
\begin{document}

\title{V, J, H and K Imaging of the Metal Rich Globular Cluster NGC\,6528
\thanks{Based on data collected at La Silla; ESO programs:
64.N-0645(A), 65.L-0610(A), and HST archival data}
}
 
\subtitle{Reddening, metallicity, and distance based on cleaned 
colour-magnitude diagrams}
\author {
Y. Momany \inst{1}
\and
S. Ortolani  \inst{1} 
\and 
E. V. Held \inst{2}
\and
B. Barbuy \inst{3}
\and 
E. Bica \inst{4}
\and
A. Renzini \inst{5}
\and
L. R. Bedin \inst{1}
\and 
R. M. Rich \inst{6}
\and 
G. Marconi \inst{7,8}
}

\offprints{Y. Momany}

\institute{
Dipartimento di Astronomia, Universit\`a di Padova,
vicolo dell'Osservatorio 2, I--35122 Padova, Italy
\and
Osservatorio Astronomico di Padova, 
vicolo dell'Osservatorio 5, I--35122 Padova, Italy
\and 
Universidade de S\~ao Paulo, Rua do Matao 1225, 05508-900 Sao Paulo, Brazil
\and 
Universidade Federal do Rio Grande do Sul, 
Dept. de Astronomia, 91501-970 Porto Alegre, Brazil 
\and
European Southern Observatory, Karl-Schwarzschild-Str. 2, D-85748
Garching, Germany
\and
UCLA, Department of Physics \& Astronomy, 8979 Math-Sciences Building, 
Los Angeles, CA 90095-1562, USA
\and
Osservatorio Astronomico di Roma, Via dell'Osservatorio 5, I-00040 
Monte Porzio, Italy 
\and 
European Southern Observatory, Alonso de Cordova 3107, Vitacura, 
Santiago, Chile
}
\date{Received 27 August 2002 /  Accepted .... 2002}
\abstract{ 
New near-infrared observations of NGC\,6528 are presented.  The
$JH$$K_{\rm s}$ observations complement a previous HST/NICMOS
data set by Ortolani et al.  (\cite{ortolani01}), in that they sample
a larger area, contain a more numerous sample of red giant stars, and
include the $K$ band.  Also, archival HST  data sets (separated by
6.093 years) were used to proper-motion decontaminate the
near-infrared sample and extract a clean $VJHK$ catalogue.
Using  the present wide   colour baseline,   we compared  the  cleaned
colour-magnitude   diagrams of NGC\,6528  with  those of NGC\,6553 and
NGC\,104 and derived new estimates of  reddening and distance, $E_{\it
B-V}=0.55$ and ($m-M$)$_{\circ}=14.44$ (7.7 kpc).
Moreover, the morphology and location of the  cleaned red giant branch
were  used to derive   a  {\it photometric}   estimate of  the cluster
metallicity.  The average of 10  metallicity indicators yields a  mean
value of [M/H] $\approx$0.0, and  [Fe/H] $\simeq  -0.20$ and $+0.08$  on
the     Zinn   \&  West (\cite{zinn84})    and     Carretta \& Gratton
(\cite{carretta97}) revised metallicity scale, respectively.
The best isochrone fit to the cleaned $K,V-K$ diagram is obtained for
a $12.6$ Gyr and Z$=0.02$ isochrone, i.e. the derived metallicity of
NGC\,6528 turns out to be very close to the mean of stars in the
Baade's Window.
%
Five AGB variable star candidates, whose membership has to be
confirmed spectroscopically, are bolometrically as bright as the known
long period variable stars in NGC\,6553.  As discussed in Guarnieri et
al. (\cite{guarnieri97}) for NGC\,6553, this may indicate that an
``intermediate age'' population is not needed to account for the
brightest stars in external galaxies such as M\,32.

}
\titlerunning{NIR and Optical Imaging of NGC\,6528}
\authorrunning{Y. Momany et al.}
\maketitle
%

%
\begin{table*}[t]
\begin{center}
\caption{Journal of observations of NGC\,6528}
\begin{tabular}{lcclccc}  
\hline\hline
\noalign{\smallskip}
Type&Filter&No. of Im.&DIT$\times$NDIT&FWHM&$X$&$\Delta m$\\
\hline
        DEEP & $J$ & 3 & 10$\times$6 & 0.78 &  1.408 & 0.080 \\ 
        DEEP & $H$ & 3 & 10$\times$6 & 0.78 & 1.366 & 0.085\\ 
        DEEP & $K_{\rm s}$ & 3 & 5$\times$12 & 0.75 & 1.328 & 0.086\\ 
        SHALLOW & $J$ & 3 & 1.18$\times$10 & 0.91 &  1.493 & 0.076\\ 
        SHALLOW & $H$ & 3 & 1.18$\times$10 & 0.91 & 1.467 & 0.081\\ 
        SHALLOW & $K_{\rm s}$ & 3 & 1.18$\times$10 & 0.85 & 1.443 & 0.079\\ 
\hline
\end{tabular} 
\label{t_log}
\end{center}
\end{table*}
\begin{figure*}
\centering
\includegraphics[width=12cm]{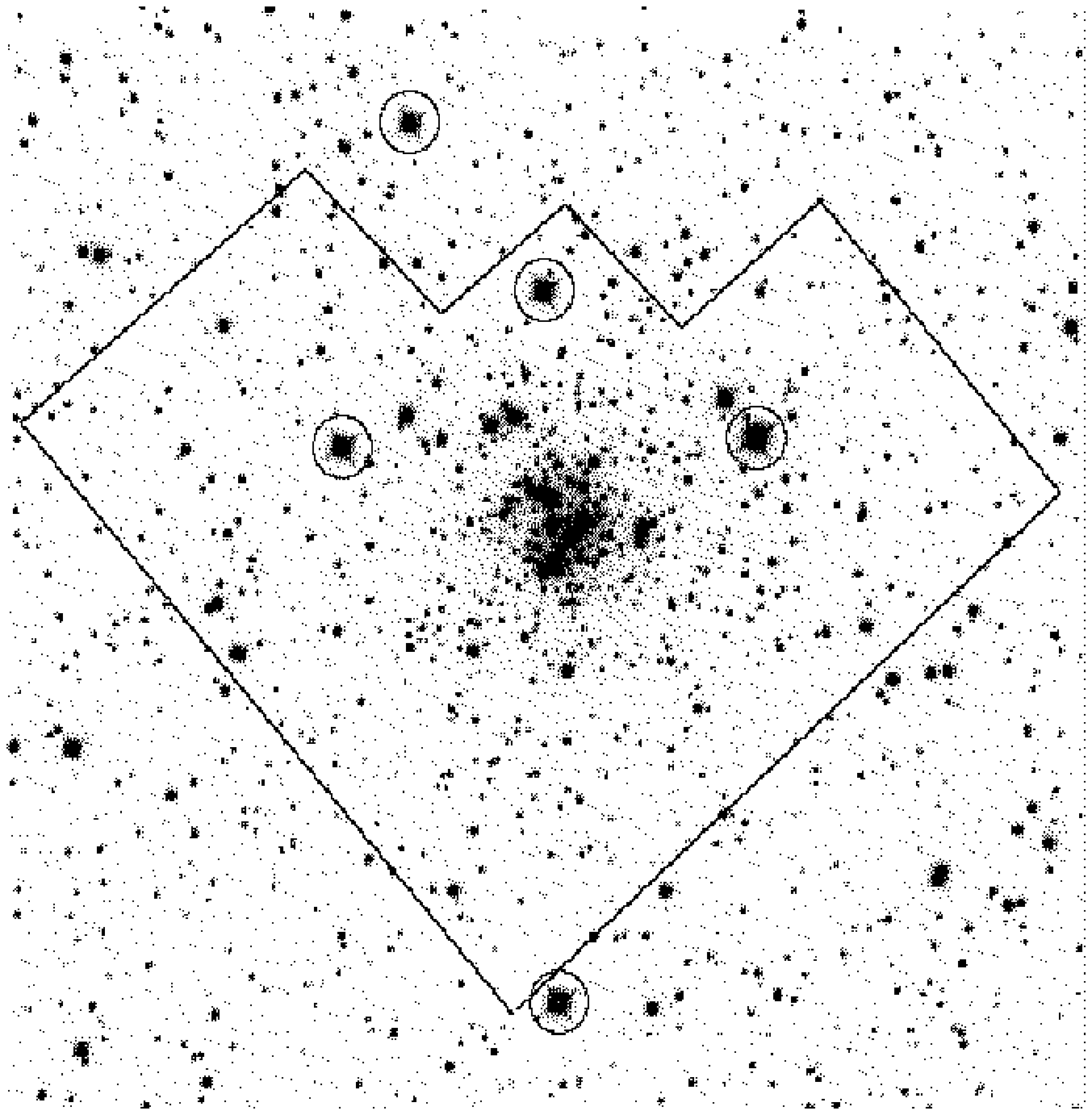}
\caption{
A $3\farcm6\times3\farcm6$ section of the {\em DEEP} 
$K$ mosaic  image of NGC\,6528.  North is to the left, 
East is to the bottom.  The field of view of
the HST observations is outlined.  Also marked are the 5 candidate
variables stars (see text). }
\label{f_fov_6528}
\end{figure*}
\section{Introduction}

The    Galactic Globular Cluster (GC) system    has long been used for
investigating  the early chemical  evolution   of the  Milky  Way.  In
particular, information  regarding the spatial distribution, dynamical
properties    and chemical abundances of     metal rich GCs  constrain
theories  on  the formation  of  the  Galactic   bulge,  hence on  the
evolution     of   the       Galaxy  as  a        whole   (Ortolani et
al.  \cite{ortolani95}).  There   are 74 globular  clusters  projected
within  20$^{\circ}$$\times$20$^{\circ}$ of the Galactic center, among
which 60 have  galactocentric distances R$_{\rm  GC}$  $<$ 4 kpc.   In
Barbuy et al.  (\cite{barbuy99}) it was shown that most of these inner
clusters  have red Horizontal  Branches (HB), whereas  the fraction of
blue   HBs increases with   the  distance  from the  Galactic  center.
Measuring  the metallicities,   colours and luminosities  of stars  in
these clusters  is hampered by many  difficulties, such as high visual
extinction, differential reddening,  crowding and  contamination  from
field stars.

\begin{table*}[t]
\begin{center}
\caption{NGC\,6528 metallicity estimates from literature}
 \begin{tabular}{llll}
\hline\hline
\noalign{\smallskip}
Method & [Fe/H] & [M/H] & Ref. \\
\hline
$Q_{39}$ index    			&	        & $+0.1$  &  Zinn \& West (\cite{zinn84}) \\
optical  diagrams 			&               & solar   &   Ortolani et  al. (\cite{ortolani92})\\
                  			&               & $-0.20$ &  Barbuy et al. (\cite{barbuy98})\\
                  			& $-$0.38$^{a}$ &         & Ferraro et al. (\cite{ferraro00})\\
IR {\rm Ca} {\rm II} integrated spectra & $-$0.23       &         & Armandroff \& Zinn (\cite{armand88})\\
IR abs. at 1.6  $\mu$m                  & $-$0.23       &         & Origlia et al. (\cite{origlia97})   \\
NIR diagrams  				&               & $+0.10$  & Cohen \& Sleeper  (\cite{cohen95})\\
IR {\rm Ca} triplet  equivalent widths  &$-$0.33        & 	  & Rutledge et  al. (\cite{rutledge97})\\
low resolution spectra 			& $-$0.50       & $-0.25$ & Coelho  et al. (\cite{coelho01})\\
high  resolution  spectra 		& +0.07         &         & Carretta    et   al. (\cite{carretta01})\\
\hline
\end{tabular} 
\label{t_lit}
\begin{list}{}{}
\item[$^{\mathrm{a}}$ on the  Carretta    \&   Gratton 
(\cite{carretta97}) scale calibration.] 
\end{list}
\end{center}
\end{table*}

Optical Colour-Magnitude Diagrams (CMDs)  of    metal rich GCs     are
characterized by the  extended turnover of  the Red Giant Branch (RGB)
caused by  TiO blanketing effects, especially  strong  in the $V$ band
(Ortolani  et  al. \cite{ortolani90},  \cite{ortolani91};   Heitsch \&
Richtler   \cite{heitsch99}).    Combining   data  from  optical   and
Near-Infrared (NIR)   allows one  to  investigate the   dependence  of
stellar properties on colour better than relying on that of optical or
NIR data alone (Cohen et al.  \cite{cohen78}; Kuchinski et al.
\cite{kuchinski95a}; Guarnieri et  al.  \cite{guarnieri98}; Ferraro et
al. \cite{ferraro00}, hereafter F00).

NIR data  have    the great  advantage  of  having  small   bolometric
corrections for cool stars, since they bracket  the spectral region of
maximum stellar flux density (Frogel et al.  \cite{frogel80}; Cohen et
al.  \cite{cohen81}). This allows  a straightforward  comparison with
theoretical models.

NGC\,6528 is   among the best studied  metal  rich GCs of  the galactic
bulge.  HST/WFPC2 observations showed that the  CMD of this cluster is
virtually identical to that of the bulge GC NGC\,6553, and that the age
of these two clusters is the same as halo GCs within a $\sim \pm20 \%$
(systematic) uncertainty (Ortolani et al. \cite{ortolani95}).

Recent distance estimates put the  cluster within $\le1$~kpc from  the
Galactic center   (Barbuy  et  al.  \cite{barbuy98}).     Both optical
(Ortolani     et        al.       \cite{ortolani95};  Richtler      et
al. \cite{richtler98}) and IR studies (Cohen \& Sleeper
\cite{cohen95}; Ferraro  et   al. \cite{ferraro00};  Ortolani et   al.
\cite{ortolani01}) agree that it is  among the most metal  rich clusters.  Although
the metal abundance estimates in the literature span a $\sim~0.6$ dex,
all values are in the metal rich regime.
This wide range in metallicity 
is partly due to uncertainties in the reddening towards NGC\,6528.
Values   in the  literature   span a very    wide range:  e.g. $E_{\it
B-V}=0.45$ (Richtler
\cite{richtler98}),    $E_{\it B-V}=0.50$  (Carretta     et   al.
\cite{carretta01}),    $E_{\it B-V}=0.52$    (Barbuy     et     al.
\cite{barbuy98}),     $E_{\it B-V}=0.55$    (Ortolani     et   al.
\cite{ortolani92}), $E_{\it B-V}=0.73$ (Reed et al.
\cite{reed88}),    up     to    $E_{\it     B-V}=0.77$  (Schlegel   et
al. \cite{schlegel98}) in the cluster direction.

Ortolani et  al.   (\cite{ortolani01}) presented HST/NICMOS  $JH$ CMDs
for NGC\,6528, reaching 3 magnitudes below the turn-off, and derived an
age of $13\pm3$ Gyr.  Their NIR diagrams are  also as ``clean'' as the
optical    diagrams  of  Feltzing et al. 
(\cite{feltzing01}),  who   used  two epochs of   HST/WFPC2  images to
proper-motion decontaminate stars  of NGC\,6528 from those belonging to
the Galactic bulge.

In this paper a  new  NIR data set   of NGC\,6528  is presented.  The
$5 \times 5$  arcmin$^2$    $JH$$K_{\rm    s}$ observations     allow the
construction of NIR CMDs that better sample the RGB and reach close to
the  turnoff. We also proper-motion   decontaminate the HST $VI$  set,
extracting a  cleaned  $VJHK$  sample.   Based on   this  wider colour
baseline  we derive new   reddening and  distance estimates, and
check  calibrations  of  photometric   metallicity
indicators.


In Sect.~\ref{s_observ} the $JHK$ and archival HST data are presented.
Section~\ref{s_reduce}  highlights  the reduction  procedures for both
data  sets.   The  original  and cleaned  NIR  CMDs  are  presented in
Sect.~\ref{s_cmd}, while in Sect.~\ref{s_hb_rgb} the  HB and RGB  bump
luminosities  are derived.   In Sect.~\ref{s_reddening} reddening  and
distance modulus   estimates  towards NGC\,6528    are  derived.    In
Sect.~\ref{s_feh} we   derive the metallicity  of  NGC\,6528  based on
various   calibrations     of     the    RGB     morphology.   Lastly,
Sect.~\ref{s_summary} summarizes our results.

\section{Observations}
\label{s_observ}
\subsection{NIR Observations}

The observations were carried out on two different runs at the ESO NTT
telescope: on 2000,  February 20$-$21 and  July 2$-$3.  Both runs used
the  SOFI  infrared    camera    equipped  with a    Hawaii     HgCdTe
$1024\times1024$ pixels   array   detector.   The  image  scale     of
$0\farcs29$ pixel  was used for all observations,  yielding a field of
view  of    $4\farcm9\times4\farcm9$.  The  read-out  mode  was Double
Correlated Read (DCR).  This yielded a readout noise  of $2.1$ ADU and
a gain of  $5.53$ electrons/ADU.  Observations of  the first  run were
conducted under  photometric  conditions  and a  seeing  of  less than
$\sim 0.8$  arcsec,   whereas the second     run  had variable  seeing
conditions, up to $1.5$   arcsec.  These observations were taken  with
two different  exposures times:  DEEP  to reach the  fainter  stars at
acceptable  S/N, and SHALLOW to  allow the  sampling of bright RGB/AGB
stars.

In Table \ref{t_log} the journal of observations is presented. Columns
1   and 2 refer to  the  type of images  taken  and  the filters used.
Column  3  (No.  of  Im) stands  for  the number  of dithered  images.
Column 4 gives the DIT   $\times$ NDIT (Detector Integration Time  and
Number of Detector Integration Time, respectively).   Columns 5, 6 and
7 give the  stellar FWHM  (in  arcsec), the airmass, and  the  derived
aperture correction.  The numbers reported in  Cols.  5, 6 and 7 refer
to  the   final  stacked    image.   Figure~\ref{f_fov_6528} shows   a
$\sim3\farcm6\times3\farcm6$ $K$  mosaic  image, corresponding   to  a
section of the entire DEEP field, upon which the  field of view of the
HST observations is outlined.

The DEEP images of NGC\,6528 were  taken in  sets of six images:
$3$  dithered  images  were  centered on  the  object  and $3$ images,
centered  in  an  adjacent  region,  were  dedicated to sky sampling. 
The dithering steps   were  of  the order of  $4\farcs0$.
Standard stars  from  Persson  et  al.  (\cite{persson98})   were also
observed on a regular basis at airmasses  comparable with those of the
target object. For each standard star five measurements were obtained:
4 measurements having the  star centered in  the $4$ quadrants of  the
detector and  one  with  the   star in the   center of the
detector.


\subsection{HST Observations}
\label{s_feltzing}

Two sets of   archive WFPC2   HST  observations, GO5436   and  GO8696,
separated by  $6.093$   years  are used.    Our   goal was to   use  a
proper-motion cleaned  $VI$   data  set  to decontaminate   our  $JHK$
photometry.  Feltzing et al. (\cite{feltzing01}) used the same data to
proper-motion  decontaminate    the optical  CMD  of    NGC\,6528.  We
therefore refer the  reader to their work  for a detailed presentation
of the two sets of observations.

\section{Data reduction}
\label{s_reduce}

\subsection{Pre-reduction}

The pre-reduction   of  the NIR    data consists  of  (1)  dark  frame
subtraction,  (2)  sky subtraction   and  (3)  flat fielding   of both
scientific   and  standard star   frames.    We basically  applied the
reduction  steps    given  in   the  SOFI   manual   (Lidman   et  al.
\cite{lidman00}) to   which  the reader is   referred   for a detailed
presentation. In   the  process   of  flatfielding    the illumination
correction frames and the bad pixels maps, both available from the ESO
webpages, were used.

A typical $J$  sequence consisted  of 3  scientific images and  3  sky
images.  We scaled the  sky images to  a common median after rejecting
the highest and lowest pixels.
On the other  hand, a typical standard
star sequence already  contains  information about its  local  sky, as
most of the  array in fact measures the  sky. Hence, at  no additional
cost  in  observing time, a median  combination  of the standard stars
sequences automatically gave their local sky frames.

Finally, both target and standard star images had their associated sky
subtracted and were divided  by their respective  flatfields. At this
point, these images were cleaned using the bad pixels maps.  To create
the combined images  of the object, scientific  images were aligned by
applying standard IRAF tasks. The matching and coaddition 
of the individual images
was facilitated  by the presence  of many  bright stellar sources, and
the  offsets were determined  accurately. 

\subsection{Calibration}
\label{S_IR_RED+CALIB}

\subsubsection{Standard Stars}

The photometric calibrations were  defined using 4 standard stars from
Persson et  al. (\cite{persson98}), namely  9106, 9119, 9143, 9172 and
the red  star LHS 2397a.   Their Table 3  showed that, for  red stars,
$K_{\rm  s}$ and $K$  are  rarely different by   more than $0.02$: the
average difference is   $0.0096$ mag,  with  a  standard  deviation of
$0.017$ mag.  We therefore   assume that the   use of the $K_{\rm  s}$
filter instead of  the $K$ filter introduces  an extra uncertainty  in
the photometric calibration of this filter of  $\pm0.02$ mag (see also
Ivanov et al.   \cite{ivanov00}).  Standard   star curves of   growth,
obtained   with the IRAF/APPHOT  task,  showed that  a  radius of $18$
pixels  (5.2  arcsec)  gave  a  satisfactory   convergence of aperture
photometry for all  the standards.  This  is consistent with the  fact
that  Persson  et al.  (\cite{persson98})  used   an aperture of  $10$
arcsec in diameter.

The study of Montegriffo  et al. (\cite{montegriffo95}) has  shown the
existence of some spatial variation in the  photometric response of IR
cameras.  We  checked   this  possibility  and   found no  significant
variations in the $5$ measurements  of standard stars.  In the process
of  calibrating the   standard  stars,  the $5$   aperture  photometry
measurements were averaged.

The instrumental magnitudes of  the  standards were normalized  to $1$s 
exposure and zero airmass, according to the following equation:

\begin{equation}
m^{'}=m_{\rm ap}+2.5\, \log(t_{\rm exp})-K_{\lambda }\, X \,
\end{equation}

\noindent 
where $m_{\rm  ap}$ is  the  mean instrumental   magnitude of the  $5$
measurements in a circular aperture  of radius $R=5.2$ arcsec, $X$  is
the  mean airmass and $t_{\rm  exp}$ is the DIT   in seconds. The mean
extinction  coefficients  adopted  for   La Silla  are:  $K_{J}=0.10$,
$K_{H}=0.04$ and  $K_{Ks}=0.05$ (from ESO  webpages).  A least squares
fit   of the normalized  instrumental magnitudes  to the magnitudes of
Persson et al.  (\cite{persson98}) gave the following relations:

\begin{equation}
J-j=+0.004\,\times (J-H)\,+\,23.089\,
\end{equation}
\begin{equation}
H-h=+0.032\,\times (J-H)\,+\,22.888\,
\end{equation}
\begin{equation}
K-k_{\rm s}=-0.001\,\times (J-K)\,-\,22.346\,
\end{equation}

The r.m.s. scatter  of the residuals of  the fit is $0.041$,  $0.020$,
and $0.021$ mag in  $J$, $H$ and $K$  respectively. Accounting for the
uncertainty in the  use of the  $K_{\rm s}$ filter  instead of the $K$
filter, we therefore assume that  $0.04$, $0.02$ and $0.03$  represent
our calibration uncertainties in $J$, $H$ and $K$ bands respectively.

\subsubsection{Calibrated Catalogues}

\begin{figure}
\centering
\includegraphics[width=8cm]{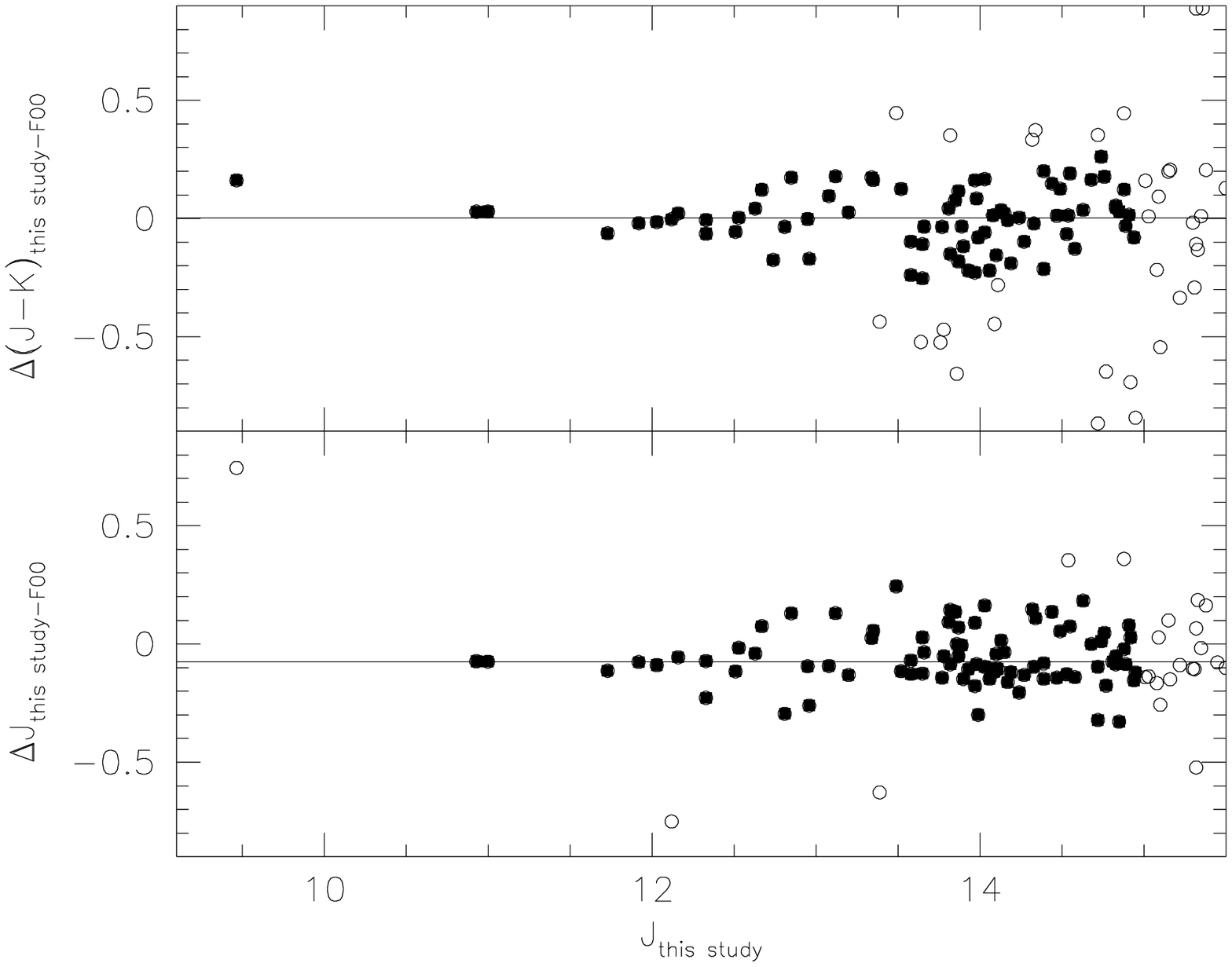}
\caption{A comparison between our $JK$ data of NGC\,6528 and those of
F00.  The horizontal line marks  the median differences after applying
a $1\sigma$ rejection (filled circles).}
\label{f_us_dav}
\end{figure}

We made use of the  DAOPHOT II and ALLSTAR (Stetson
\cite{stetson87}, \cite{stetson94})  packages for  stellar photometry.
After running the FIND and PHOT  tasks, we searched for isolated stars
to build the PSF for  each of the  final images.  These PSF candidates
showed  no faint  components   and the  subtraction  turned  out to be
smooth.  The final PSF were generated with a PENNY function that had a
quadratic dependence on  position in the frame. The  photometry
was finally performed on the stacked images using ALLSTAR.

The instrumental photometric catalogues contain PSF magnitudes.  These
were converted    into  aperture  magnitudes   assuming that   $m_{\rm
ap}=m_{\rm  PSF}-constant$  (Stetson   \cite{stetson87}),   where  the
constant   is  the  aperture correction to   be   found.  The aperture
corrections were estimated  as in the   optical case: bright  isolated
stars  were selected  in each  field and  had their neighbours (within
$5.8$   arcsec  radius) subtracted,    then aperture  magnitudes  were
measured through  increasing  circular   apertures with  radii  within
$R=3.5-5.8$  arcsec.   The curves of   growth  of the  brightest stars
showed a  satisfactory aperture photometry  convergence at a radius of
$18$ pixels ($5.2$ arcsec), the same  as for the  standard stars.  The
aperture corrections are  reported  in Col.  7   of Table~\ref{t_log}.
Having estimated the aperture  corrections, calibrated magnitudes were
obtained by    applying Eqs. 2--4,  and  the   final DEEP and  SHALLOW
catalogues of NGC 6528 were produced.

All catalogues in the same  passband were then matched.  For stars in
common, we computed     the  error-weighted mean of   the    magnitude
differences.   Note that the  SHALLOW and  DEEP  catalogues  (for each
passband) overlap  over an interval of $\sim  5$  magnitudes, and that
the magnitude  differences did not  exceed $0.03$ mag.  At this point,
the  mean of  the magnitude differences   was  applied to the  SHALLOW
catalogues so as to  share the same magnitude scale  of the DEEP ones.
Stars found {\it only } in either  catalogues are flagged and included
anyhow in the final catalogue, whereas those found in both SHALLOW and
DEEP were  recorded with their  DEEP  magnitudes, that have  a  higher
signal to  noise.  Finally, a  $JHK$ colour  catalogue  is produced by
associating   the entries of  the  combined single passband catalogues
described above.

\subsection{Comparison with Previous Photometry}

Ferraro et al.  (\cite{ferraro00}) presented a homogenous NIR database
of 10 Galactic  GCs, obtained at the  ESO/MPI~2.2 m telescope equipped
with the  NIR  camera IRAC-2.  A  bright  star sample  of NGC\,6528 was
matched with the photometry of Ferraro et al.,  kindly provided by the
authors.   The panels of  Fig.~\ref{f_us_dav}  show  the results of  a
comparison of our photometry with  the Ferraro et al. sample; applying
a  $1\sigma$ rejection yielded  the  following values:  $\Delta J_{\rm
this~study-F00}=-0.076\pm0.110$        and       $\Delta$($J-K$)$_{\rm
this~study-F00}=0.003\pm0.120$.   The  differences in  the $J$ and $K$
scales  are  of  the order    of the  uncertainties   of the  aperture
corrections, and  overall show  the excellent  consistency  of the two
magnitude scales.

\subsection{HST reduction and proper-motion decontamination}
\label{s_deco}

The methods used to  reduce  the HST  WFPC2 archive  observations  are
similar to those employed by  Anderson \& King (\cite{anderson00}) and
will not be  further commented here.  Instead,  we briefly comment  on
the methods used  to derive a cleaned CMD.
To  carry out  the  astrometry, the   algorithms of  Anderson \&  King
(\cite{anderson00}) based on the {\em effective} point-spread function
(PSF)  were used.  The basis  of the method is   to determine a finely
sampled PSF of  high accuracy  from  images at dithered  offsets.  The
fitting of this PSF  to individual images  gives a positional accuracy
of the order of $\sim0.02$ pixels, showing no systematic error arising
from the location of the star with respect to the pixel boundaries.
Since all astrometric measurements were made with respect to reference
stars that are  cluster  members,  the  zero  point of  motion is  the
centroid  motion of the cluster (King  et  al. \cite{king98}; Bedin et
al. \cite{bedin01}).

It is important to note that for stars brighter than $V\sim16.8$
($K\sim12$), the  core of the PSF  is flattened due  to saturation, and
this is where most of the astrometric  information is contained (WFPC2
is  severely undersampled).   Therefore, any tentative decontamination
of  cluster  stars   brighter  than  ($K\sim12$)     is hampered by
saturation.  Besides saturations effects,   many bright stars  in  our
$JHK$ sample are not  included in the decontamination  process because
they lie outside   the   HST field  of    view or  do not  have   $VI$
counterparts within 1 pixel matching radius.

The decontamination process resulted  in $VI$ diagrams, that are  very
similar   to those   obtained by  Feltzing et al. 
(\cite{feltzing01}). We therefore use their same cuts on the retrieved
stellar  proper   motion  in   the  $l$   and  $b$  coordinates,  i.e.
$\sqrt{\mu_{l}^2+\mu_{b}^2}<0\farcs09$      for     $V\ge19$       and
$\sqrt{\mu_{l}^2+\mu_{b}^2}<0\farcs23$     for $V < 19$.  Lastly,  the
decontaminated $VI$ data  were used to extract  a {\it cleaned} $VJHK$
sample of NGC\,6528 stars. For the rest of  this paper by {\it cleaned}
diagram we refer to the product of imposing the above cuts.

\section{The Colour-Magnitude Diagrams}
\label{s_cmd}

\begin{figure}
\centering
\includegraphics[width=8cm]{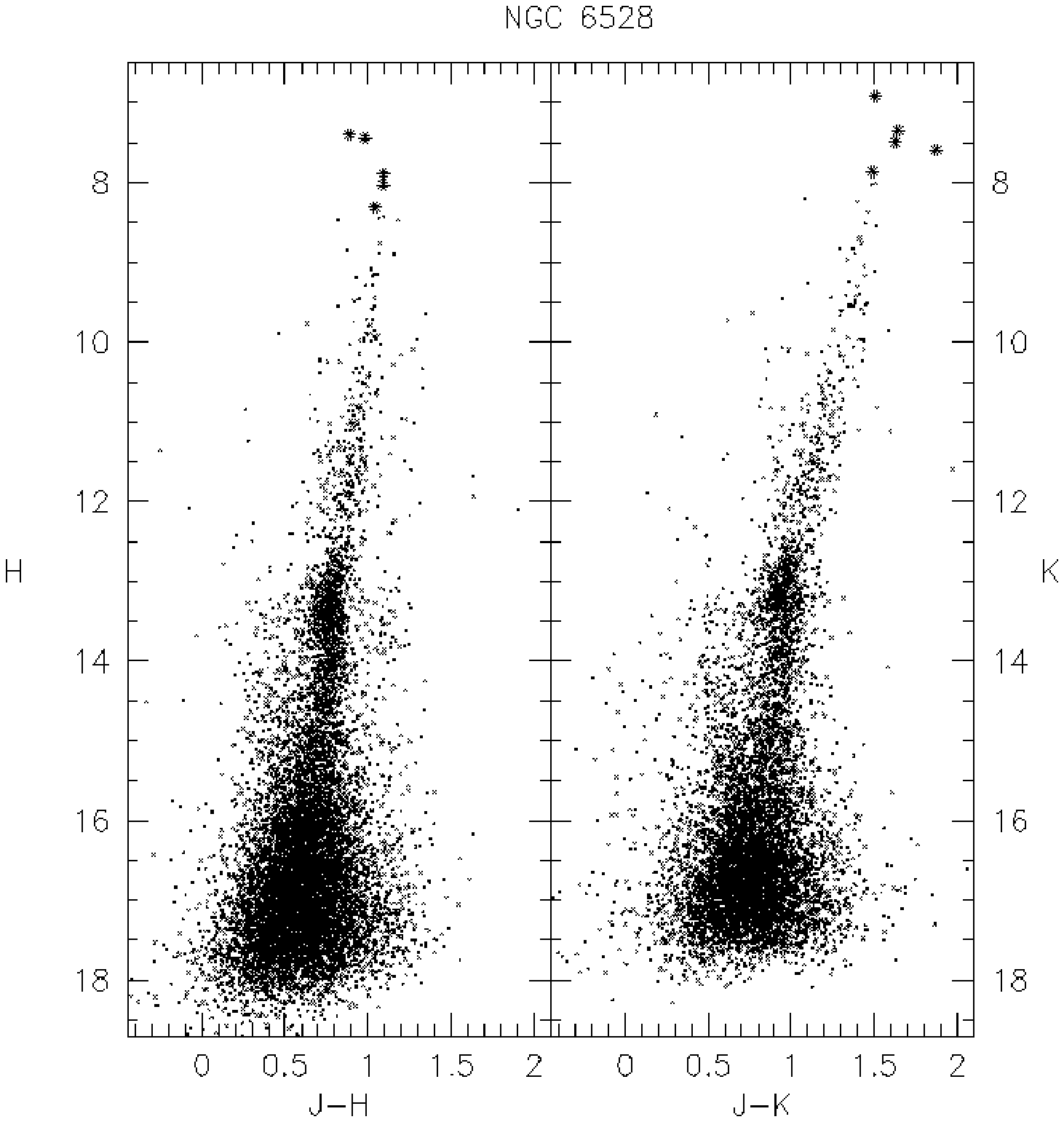}
\caption{The original $H, J-H$ and $K, J-K$  
colour-magnitude diagrams of NGC\,6528.  Only stars with DAOPHOT errors
less than $0.08$ have been plotted.  Starred symbols are the candidate
variable stars (see text).}
\label{f_cmds1}
\end{figure}

Figure~\ref{f_cmds1} presents  the $H, J-H$  and $K, J-K$  diagrams of
the  observed field towards NGC\,6528   (only stars with DAOPHOT errors
less than $0.08$ mag  have been plotted).  This figure shows the CMDs
originating from the whole $5\times5$ arcmin$^2$ SOFI field.  The wide
cluster coverage  has the side effect of  high contamination  by bulge
stars that outnumber those of the cluster.  Note that NGC\,6528 is well
inside the Galactic  bulge and its metallicity  
(listed in Table 2) is  close to
the   mean   value  for bulge stars ([Fe/H]$=-0.25$;    McWilliam   \&  Rich
\cite{mcwilliam94}, hereafter MR94).  Therefore  one expects the  main
features  of the CMD  (RGB,  HB and MS) to  be  populated by both  the
cluster and bulge populations (in addition to the disk contamination).
On the other hand, the diagrams  show a well sampled, almost vertical,
RGB sequence  spanning from  $K\simeq16.5$  to  $K\simeq7.0$.   The
present photometry is also deep enough to reach the cluster's subgiant
branch.

\begin{figure}
\centering
\includegraphics[width=8cm]{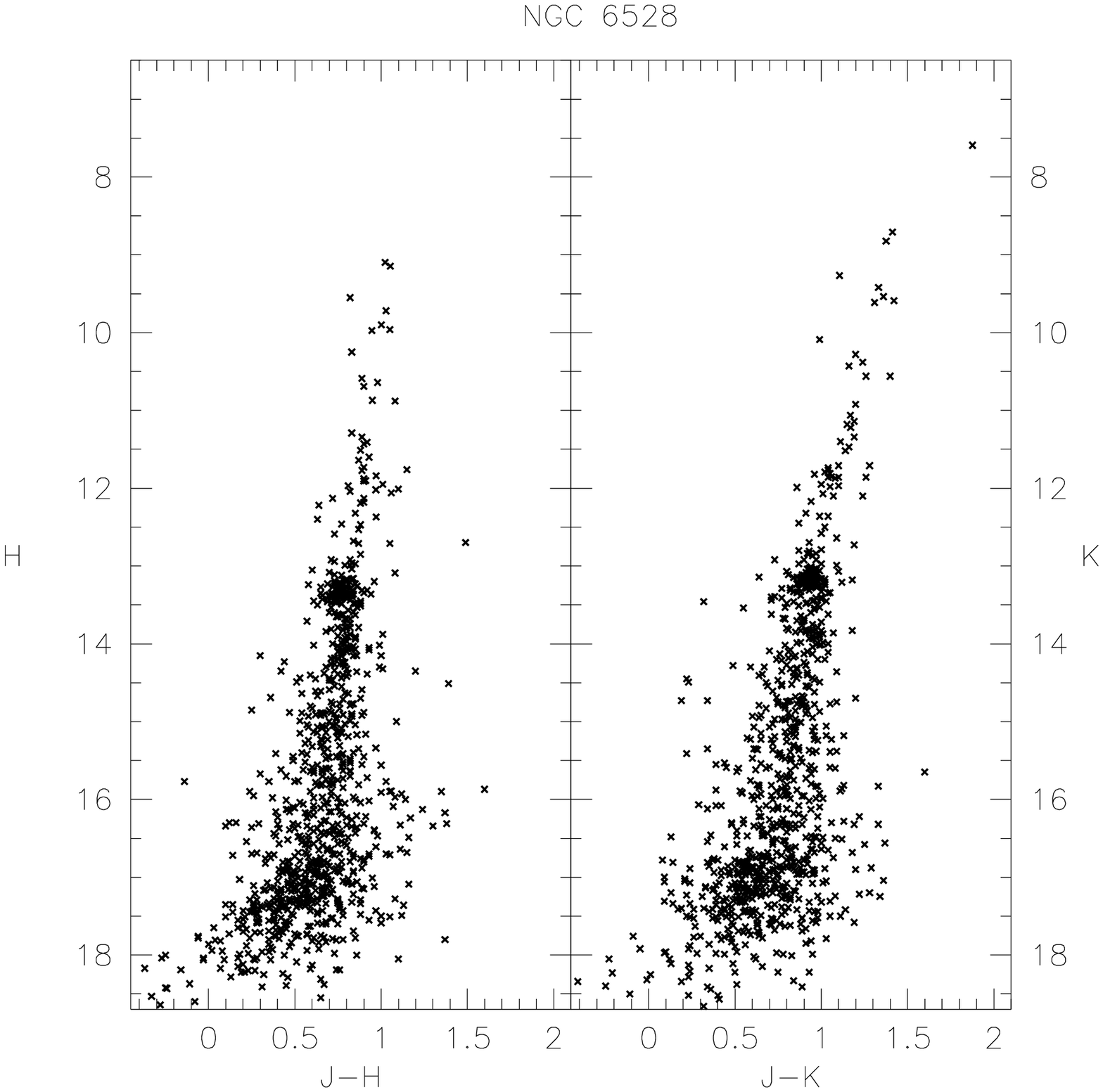}
\caption{As above but for the decontaminated diagrams.}
\label{f_cmds2}
\end{figure}

Figure~\ref{f_cmds2} displays  the corresponding decontaminated  CMDs,
which  refer  to  the much  smaller  HST  field where the  astrometric
decontamination  was conducted.  Hence, the decrease  in the number of
RGB stars in the  lower panels is  not only due to the decontaminating
process but also to limitation of the WFPC2 field of view. The cleaned
$JHK$ data, and the subsequent matching with the  HST $V$ data provide
the basis for our further analysis of the cluster's parameters.

\begin{figure}
\centering
\includegraphics[width=8cm]{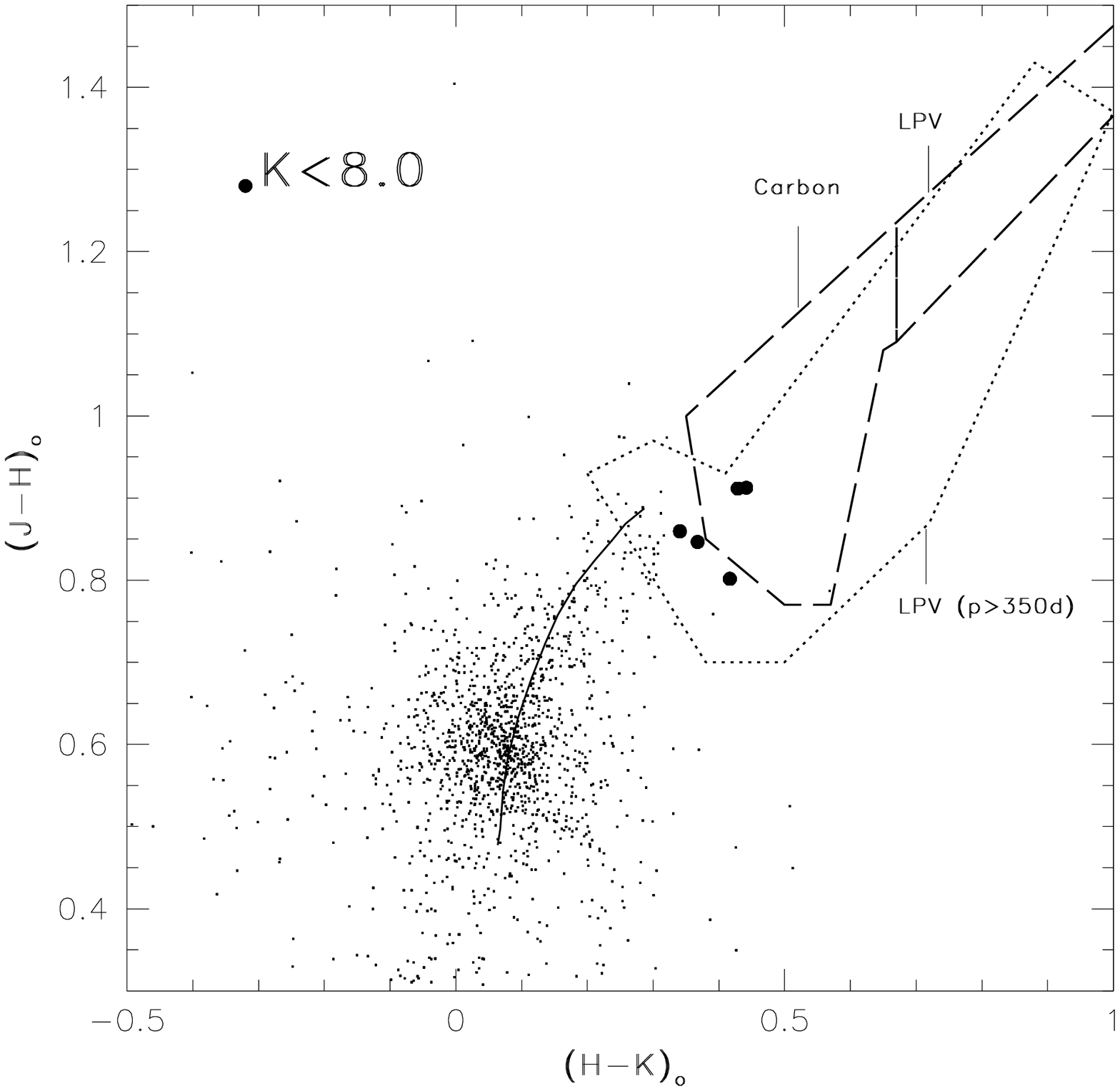}
\caption{The dereddened $(J-H)_{\circ}$, $(H-K)_{\circ}$  two colour diagram of NGC\,6528.  
The overplotted loci are those  by Bessell \& Brett (\cite{bessell88})
for carbon  stars and long period  variables (long-dashed), and Frogel
et al.  (\cite{frogel78}) for variables with  $P>350$ d (dotted).  The
mean locus for  K and M giants by  Bessell \& Brett (\cite{bessell88})
is also plotted (solid line).}
\label{f_cmds_bessell}
\end{figure}

In   Fig.~\ref{f_cmds_bessell}     we   show the   dereddened    (see
Sect.~\ref{s_reddening}  for the    adopted  reddening and   distance)
$(J-H)_{\circ}$,  $(H-K)_{\circ}$ two colour diagram.  The overplotted
loci are those by Bessell \& Brett (\cite{bessell88}) for carbon stars
and     long period  variables    (long-dashed), and    Frogel et  al.
(\cite{frogel78})  for variables  with  $P>350$ d  (dotted).  The mean
locus for K and  M giants  by  Bessell \& Brett (\cite{bessell88})  is
also plotted (solid line).  This figure clearly  shows that most stars
in our sample are  K and M giants.  Moreover,  that a number  of stars
have IR  colours typical  of carbon and   long period variable  stars.
Among   these   we plot   as   filled   circles  stars  with   $K<8.0$
(M$_{bol}^{tip}\simeq-3.6$), and cautiously argue  that these have the
highest probability  of   being AGB stars   (also   plotted as starred
symbols in Figs.~\ref{f_cmds1} and \ref{f_cmds_bessell}).  As shown by
Frogel  et    al.   (\cite{frogel90}),   all   stars  brighter    than
M$_{bol}=-3.6$ are  exclusively  AGB objects, while fainter  than this
limit they  are  mostly RGB  objects.  This  interpretation is in good
agreement   with  the estimated    RGB   tip of  NGC\,6528  being   at
M$_{bol}^{tip}=-3.68$ (F00).  However, in the absence of spectroscopic
confirmation     and       given the    difficulties     explained  in
Sect.~\ref{s_deco}, we can not firmly establish the cluster membership
of these variable candidates.

\begin{figure}[t]
\centering
\includegraphics[width=8cm]{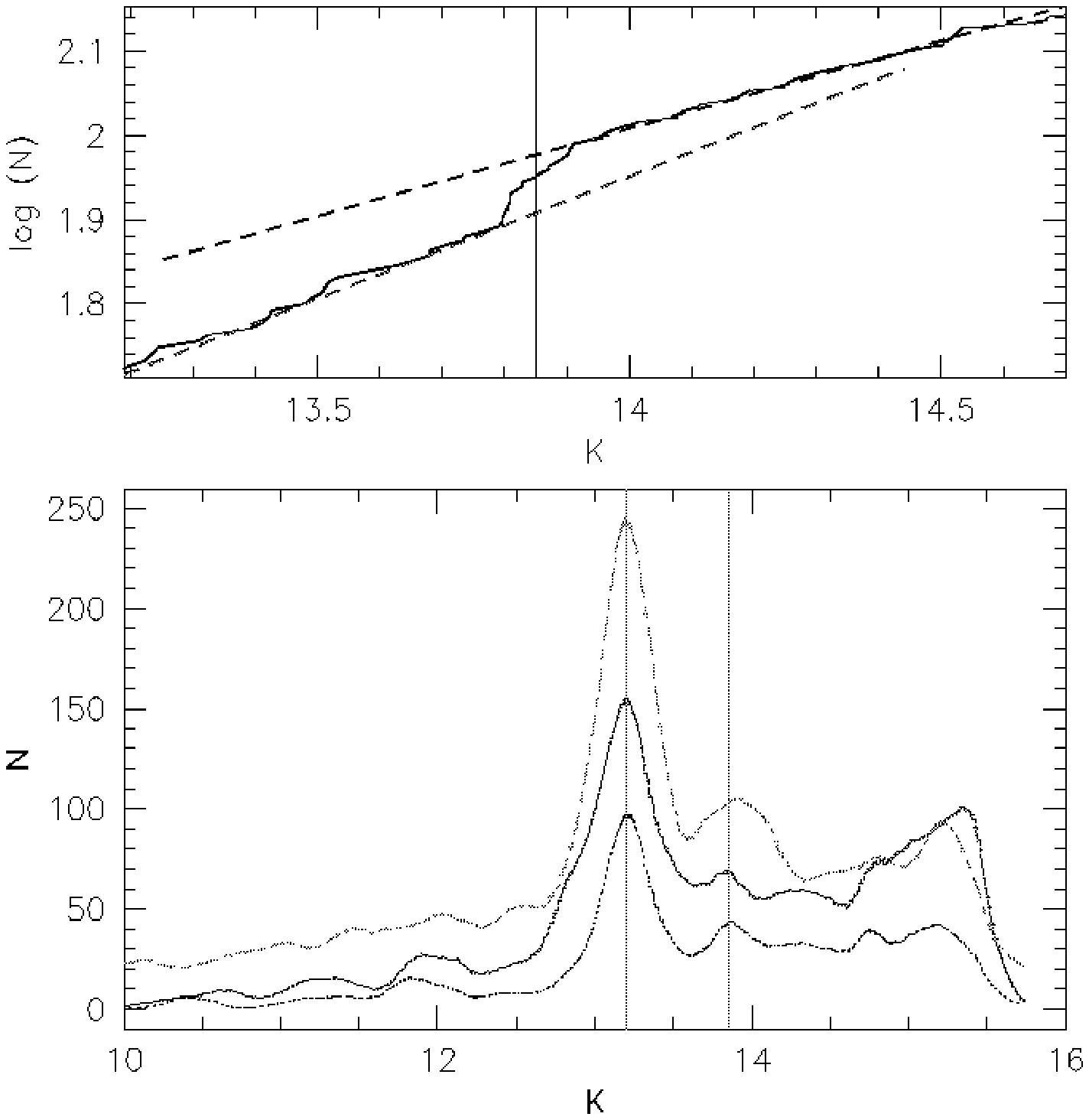}
\caption{ The upper panel shows the cumulative  logarithmic $K$ 
luminosity  function of a cleaned RGB  sample. The dashed lines
are linear fits to the regions above and  below our estimated location
of  the bump, marked by  the vertical line.  The  lower panel shows a
multi-bin,  multi-interval, smoothed differential  LF of (1) NGC\,6528
original data (solid line); (2)  NGC\,6528 {\it cleaned} data  (dotted
line); and (3)  NGC\,6553 original data  (dashed line) by Guarnieri et
al.  (\cite{guarnieri98}).    Solid vertical lines  mark the estimated
location of the RGB bump and HB in NGC\,6528.}
\label{f_hb_bump}
\end{figure}

\section{The HB and the RGB bump}
\label{s_hb_rgb}

The   brightness    and amplitude of  the   RGB   bump are metallicity
dependent,  becoming  stronger and  fainter as   metallicity increases
(F00).  Therefore the  identification of the RGB bump  can be used  to
derive  relative metallicities.    Following the  methods described in
Fusi Pecci   et    al.    (\cite{fusi90}),   the upper    panel     of
Fig.~\ref{f_hb_bump} shows  the  cumulative logarithmic $K$ luminosity
function (LF)  of a decontaminated RGB sample.   The dashed  lines are
linear fits to the regions above and below  the bump and highlight the
change in the slope of the RGB cumulative LF.  The vertical line marks
the    RGB     bump  that   we    identify    at   $K_{\rm  Bump}^{\rm
NGC\,6528}=13.85\pm0.05$.  This value is  $0.2$ mag brighter than that
found  by F00,  and  in good agreement  with   the value of  $K=13.80$
derived by Davidge (\cite{davidge00}).

The lower  panel   of   Fig.~\ref{f_hb_bump} presents   3   multi-bin,
multi-interval, smoothed differential  LFs of (1)  NGC\,6528, original
data (solid line);  (2) NGC\,6528, {\it  cleaned} data  (dotted line);
and (3) NGC\,6553,  original data (dashed  line)  by Guarnieri et  al.
(\cite{guarnieri98}), shifted  to   the  distance   and  reddening  of
NGC\,6528.  Following Bedin et al. (\cite{bedin00}), these curves were
constructed   in   a  way  to  reduce   the   effects of   statistical
fluctuations,  the  assumed interval   in  the $K$ magnitudes,  or the
assumed bin size. We first allowed the bin size to vary between $0.10$
and $0.50$ in steps of $0.01$ mag. Then, and for  each of the 40 bins,
different LFs were constructed  by changing the starting $K$ magnitude
interval.  The distribution  of these LFs  (sharing the same bin  size
but  not the same $K$  interval) was used to assign  a {\it single} LF
for that particular bin size.  Finally, the 40 LFs were normalized and
smoothed.

The most conspicuous  feature  in  Fig.~\ref{f_hb_bump} is   the  peak
marking the HB level. The  {\it  cleaned} and original NGC\,6528  data
show an excellent agreement on the HB  being at $K=13.20\pm0.05$.  The
second  notable feature is the  RGB bump.  The {\it cleaned} NGC\,6528
curve shows this feature at  $K=13.85$, in perfect agreement with that
derived from the cumulative  LF.  Also, Fig.~\ref{f_hb_bump} shows how
NGC\,6528 and NGC\,6553 share essentially  the same separation between
the HB  and  RGB bump levels.   This confirms  earlier evidence of the
similar metallicity of the two objects (Ortolani et al.
\cite{ortolani95}).

\begin{figure}[t]
\centering
\includegraphics[width=8cm]{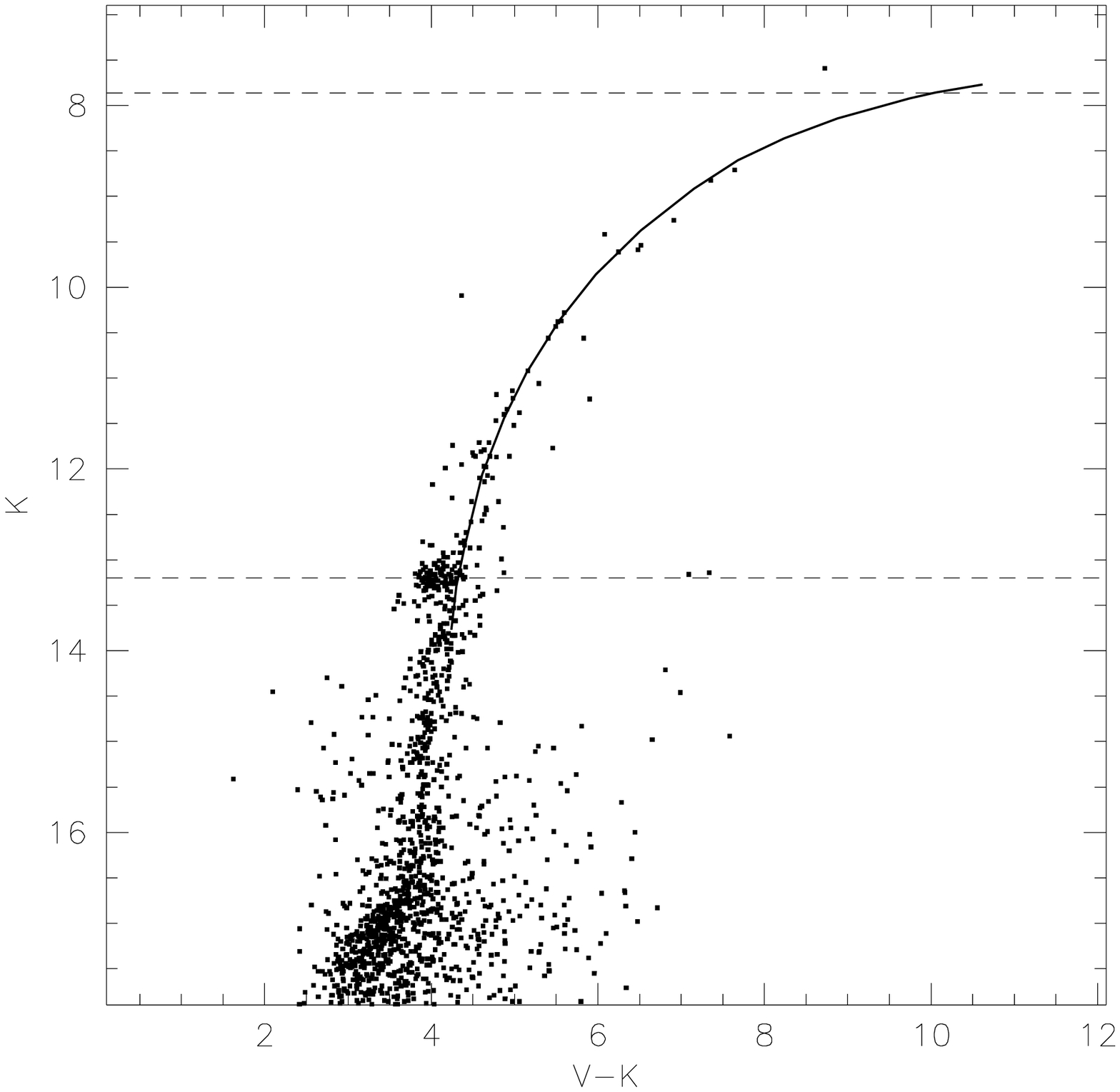}
\caption{The {\it cleaned} $K,V-K$ diagram of NGC\,6528 along with the
shifted  fiducial line   (solid  curve)   of NGC\,6553  (Guarnieri   et
al. \cite{guarnieri98}).  The dashed lines mark the HB and the RGB tip
of  NGC\,6553,  which coincide  perfectly   with their  counterparts in
NGC\,6528.}
\label{f_vkk_6553}
\end{figure}

\section{Reddening \& distance}
\label{s_reddening}

The interstellar reddening and distance of NGC\,6528 was estimated by
comparing its CMDs with the fiducial lines of NGC\,6553 in various
planes.  Also, a similar comparison was done with respect to NGC\,104
(47~Tuc).  The CMDs of NGC\,6528 are, in fact, almost identical to those
of NGC\,6553 (Ortolani et al.  \cite{ortolani95}).  This is best shown
in Fig.~\ref{f_vkk_6553} which displays the {\it cleaned} $K,V-K$
diagram of NGC\,6528 along with the shifted fiducial line of NGC\,6553
(Guarnieri et al.  \cite{guarnieri98}).
The dashed lines mark the HB and RGB tip of NGC\,6553
from Guarnieri et  al. (\cite{guarnieri98}).

\begin{table}
\begin{center}
\caption{Summary of the comparison between the CMDs of NGC\,6553 and NGC\,104 with respect to NGC\,6528}
\begin{tabular}{lllll} 
\hline\hline
\noalign{\smallskip}
NGC    & $\Delta$  colour   &   $\Delta$   mag   &  $E_{\it  B-V}$   &
($m-M$)$_{\circ}$ \\
\hline
6553 & $\Delta  (J-K)=-0.03$ & $\Delta   K=0.78$  & 0.64 & 14.38  \\
6553 & $\Delta  (V-K)=-0.35$ & $\Delta K=0.78$  &  0.57 & 14.38  \\
104 & $\Delta (J-K)=+0.36$ & $\Delta J=1.20$ & 0.44 & 14.55 \\
\hline
\end{tabular} 
\label{t_redd}
\end{center}
\end{table}


The results of the comparison between the CMDs of NGC\,6528 and those of 
NGC\,6553 and NGC\,104 are reported  in Table~\ref{t_redd}, in which
Col. 2
and 3  refer to colour and  magnitude shifts applied to the comparison
cluster in order to match  the CMD of  NGC\,6528.  The last two columns
report the derived values of $E_{\it  B-V}$ and ($m-M$)$_{\circ}$.  In
this comparison we assumed:

\begin {itemize}
\item $E_{\it B-V}=0.70$  and  ($m-M$)$_{\circ}=13.60$ for NGC\,6553 (Guarnieri  et al. \cite{guarnieri98});
\item $E_{\it B-V}=0.04$,  ($m-M$)$_{\circ}=13.32$,
      and [Fe/H]$=-0.71$ for NGC\,104 (F00);
\item $A_{\it V}=3.1\times E_{\it B-V}$;
\item the reddening law of Rieke \& Lebofsky (\cite{rieke85}).
\end{itemize}

When shifting the RGB  fiducial line of NGC\,104 with respect
to NGC\,6528, one must keep in mind that part of the colour shift is due
to  the different metallicities of  the two clusters.   We accounted for
this effect  by  using two   isochrones from   the Marigo  \&  Girardi
(\cite{marigo01}) library to estimate  the $J-K$ colour difference of
the two isochrones at the HB level.  This value  was found to be $\sim
0.11$  mag,  and  was subtracted   from   the $J-K$  colour difference
reported in Table~\ref{t_redd}.

The mean of the   reddening values reported in  Table~\ref{t_redd}  is
$E_{\it B-V}^{\rm NGC6528}=0.55$, and we will adopt this value for the
rest of the paper.  This reddening value is intermediate when compared
to the    values  reported in the  literature,   ranging from $E_{\it
B-V}=0.45$   (Richtler et   al.   \cite{richtler98})    up  to $E_{\it
B-V}=0.77$ (Schlegel et al. \cite{schlegel98}).  It is also consistent
with that   inferred  from Carretta  et al.    (\cite{carretta01}) who
estimate a mean excitation  equilibrium temperature of  4 red-HB stars
to be $T_{\rm eff}\sim4610$ K,  yielding to $E_{\it B-V}=0.50$.   Note
however that excitation   temperatures suffer  deviations  from  local
thermodynamic   equilibrium (NLTE)  effects on   FeI  lines, and  give
systematically higher values than photometric temperature (Th\'evenin
\& Idiart    \cite{thev99}), therefore this  does   not give  a strong
constraint on the reddening value.  A discussion  on the $E_{\it B-V}$
of  NGC\,6528 using integrated   spectra  is given in  Bruzual   et al.
(\cite{bruzual97}).      From  the  spectral  energy  distribution  of
NGC\,6528, they derive $E_{\it B-V} = 0.62$.

The  derived mean        distance   modulus of NGC\,6528 is
($m-M$)$_{\circ}=14.44$  (7.7  kpc),   i.e.  $0.05$ larger  than  that
obtained by Ortolani et al.  (\cite{ortolani92}).  
An independent check on the mean distance modulus was obtained from
the $K$ mean luminosity of the HB level, using the Guarnieri's et
al. (\cite{guarnieri98}) calibration. Guarnieri et al.
(\cite{guarnieri98}) observed six globular clusters with metallicities
in the range $-2.3 <$[Fe/H]$<-0.14$, and determined $M_{K}^{\rm HB} =
-0.20 \mbox{[Fe/H]} -1.53$, where [Fe/H] is on the Zinn \& West
(\cite{zinn84}, hereafter ZW) scale. Assuming [Fe/H]$^{\rm
NGC6528}_{\rm ZW}=-0.20$ (see Sect.~\ref{s_otherZ}), we derive $M_{\it
K}^{\rm HB}=-1.49$.
From the measured mean level of the HB, $K_{\rm HB}=13.20$, and
adopting an extinction $A_K=0.193$, we theferore obtain a distance
modulus ($m-M$)$_{\circ}=14.50$.  This value is only $0.06$ mag larger
than the value obtained from the CMD fitting.



\section{Metallicity}
\label{s_feh}

In Table~\ref{t_lit}   we  report   literature  metallicity   values
assigned to  NGC 6528.   In discussions about   metallicity values, it is
important  to  distinguish  between [Fe/H]   and   [M/H].    Strictly,
[Fe/H]=log(Fe/H)$_*-$ log(Fe/H)$_{\odot}$  is the  iron abundance,
as commonly derived  from the numerous Fe I  and Fe II lines available
in high resolution spectra.   [M/H] gives the  abundance of metals; if
some elements are in excess relative to  Fe, [M/H] has to  be
computed taking into account these excesses. Z  is further weighted by
the  mass  of the  elements,  but in  [Z]  (more usually  indicated by
[Z/Z$_{\odot}$]) the masses  are  cancelled, such  that  [Z] =  [M/H].
Isochrones are calibrated in terms of Z.
Ultimately, both photometric measurements and isochrone determination
depend on a calibration to be given in terms of [Fe/H], which is the
primary calibrator (since it is directly measured from spectra).

%
The RGB  morphology  in the  CMDs of globular  clusters is  a powerful
metallicity indicator, due to its  strong dependence on continuum  and
line  blanketing by heavy  elements, and essentially  no dependence on
helium abundance and age.  The behaviour of the  RGB morphology in the
optical was discussed by Ortolani et  al.  (\cite{ortolani91}), and in
the  infrared by Kuchinski et  al.  (\cite{kuchinski95a}) and F00.  It
is  worth mentioning that   the slope in the  $K,V-K$  CMD is far more
sensitive to metallicity than in a fully IR CMD (e.g.   $K, J-K$).  In
the   following  section, we   analyze   a  number of  features   that
characterize   the  RGB morphology    and  are  used  as   photometric
metallicity indicators.

%

\subsection{RGB slope}
\label{s_slope}

In the $K,  J-K$  plane,  Kuchinski \&  Frogel   (\cite{kuchinski95b})
investigated the behaviour of the RGB slope  in metal rich GCs ($-$1.0
$<$ [Fe/H] $<$ $-$0.3), and found that  the slope ($S_{\rm RGB}$) of a
linear fit to  RGB stars brighter than the  HB can be used to estimate
the metallicity  with an accuracy  of $\pm$0.25 dex. The importance of
this metallicity indicator relies on the fact that it is reddening and
distance independent.   Following their indications, we selected stars
from 0.6 to 5.1 mag above the HB level on the {\it cleaned} $K$, $J-K$
diagram of   NGC 6528.  Since  the method  is very sensitive  to small
variations in the  adopted   slope,    special care  was  needed    in
eliminating cluster HB stars and field stars. We found that a variation
of  0.01 in  the slope  translates   into  a change in  the  estimated
metallicity of  about $\sim$0.2  dex.  We  derive  a slope of $-$0.109
from   the   {\it  cleaned}   CMDs.  Using  the   Kuchinski  \& Frogel
(\cite{kuchinski95b})   calibration this corresponds to [Fe/H]$=-$0.38
on the ZW scale.


\subsection{Other [Fe/H] indicators}
\label{s_otherZ}

Ferraro et al. (\cite{ferraro00}) presented a homogeneous NIR database
of 10 globular clusters spanning a  wide metallicity range.  Analyzing
the RGB  morphology   they  calibrated a  variety  of  observables  as
metallicity  indicators (RGB slope,  RGB   $J-K$ and $V-K$ colours  at
different  magnitude  levels,  and RGB  $K$   magnitudes at  different
colours), and  based their calibrations   on  the Carretta \&  Gratton
(\cite{carretta97}, hereafter G97) metallicity scale.  Columns 1 and 2
in Table~\ref{t_ferraro} report the applied F00 calibrations and their
corresponding [Fe/H]$_{\rm CG97}$  estimates.  All  these values  were
obtained on  the  the {\it cleaned} $J-K$,  $V-K$  and  M$_{\rm bol}$,
log($T_{\rm  eff}$)  diagrams.  Besides,  we   use the  Cho  \&  Lee's
(\cite{cho02})\footnote{As   pointed    out         by       Carpenter
(\cite{carpenter01}), the Persson et al.  (\cite{persson98}) standards
(used in this  paper) have been adopted  as the fiducial calibrator in
29 of the  35 2MASS calibration  fields,  thereby connecting  the zero
points of the 2MASS to  the Las Campanas Observatory (LCO) photometric
system.  Hence, a roughly zero  photometric offset is expected in  the
2MASS-LCO transformation equations.   This is confirmed by Carpenter's
equation 22.}  calibration of the RGB bump, identified in  11 GCs (based on
data from the 2MASS point source catalogue).

%
%

\begin{table}
\begin{center}
\caption{Metallicity estimates based on the F00 
calibrations of RGB slope, and RGB specific colours and magnitudes}
\begin{tabular}{llll} 
\hline\hline
\noalign{\smallskip}
Metallicity indicator &[Fe/H]$_{\rm CG97}$&$^{a}$[Fe/H]$_{\rm ZW}$&$^{b}$[Fe/H]$_{\rm CCG0B1}$\\
\hline
$M_{K}^{(J-K)_{\circ}=0.7}$      &  $-0.10$ &  $-0.12$  &$+0.27$\\
$M_{K}^{(V-K)_{\circ}=3.0}$      &  $+0.01$ &  $-0.19$  &$+0.10$\\
$(J-K)_{\circ}^{M_K=-5.5}$       &  $-0.04$ &  $-0.17$  &$+0.15$\\
$(J-K)_{\circ}^{M_K=-5.0}$       &  $+0.02$ &  $-0.14$  &$+0.22$\\
$(V-K)_{\circ}^{M_K=-5.5}$       &  $-0.29$ &  $-0.27$  &$-0.08$\\
$(V-K)_{\circ}^{M_K=-5.0}$       &  $-0.27$ &  $-0.23$  &$+0.01$\\  
Log   $T_{\rm eff}^{\rm  bump}$  &  $-0.17$ &  $-0.19$  &$+0.10$\\  
$M^{\rm   bump}_{\rm bol}$       &  $-0.14$ &  $-0.18$  &$+0.12$\\ 
RGB slope                        &  $-0.42$ &  $-0.18$  &$+0.12$\\
\hline				     	          
$^{c}$RGB slope                     & $-0.38$  & $-0.32$&$-0.18$\\
\hline				     	          
$^{d}$$M_K^{\rm Bump}$              & $-0.42$  &        &\\
\hline
\end{tabular} 
\label{t_ferraro}
\begin{list}{}{}
\item[$^{\mathrm{a}}$ our calibration of the F00 data on the ZW scale]
\item[$^{\mathrm{b}}$ values in Col. 3 transformed onto the CCGB01 scale] 
\item[$^{\mathrm{c}}$ based on the Kuchinski \& Frogel
(\cite{kuchinski95b}) calibration]
\item[$^{\mathrm{d}}$ based on the Cho \& Lee (\cite{cho02}) calibration] 
\end{list}
\end{center}
\end{table}
%

The CG97 scale, however, has been revised by Carretta et al.
(\cite{carretta01}, CCGB01) based on high-dispersion spectroscopy of
NGC\,6528 and NGC\,6553.  In order to provide metallicity estimates on
the new CCGB01 scale and on the widely used scale of ZW, we use the
tables of F00 to (1) re-calculate the F00 calibrations on the ZW scale
(Col.  3 in Table~\ref{t_ferraro}), and (2) place these latter values
onto the CCGB01 scale (Col.  4), by applying equation 3 in Carretta et
al.  (\cite{carretta01}).  Averaging the values of Col.  3 yields 
[Fe/H]$_{\rm ZW}=-0.20$, while the average of Col.  4 gives
[Fe/H]$_{\rm CCGB01}=+0.08$, and we will adopt these values as our best
estimates.  Similarly, by applying various F00 calibrations of the
global metallicity, we derive a mean value  [M/H] $\approx$ 0.0.
Indeed, the expected [M/H] value should be $\sim 0.2$ dex higher than
[Fe/H] if the overabundance of $\alpha$-elements in NGC\,6528 is
similar to that of bulge stars (e.g., MR94).


%

%
%

\subsection{Isochrone fitting}

\begin{figure}
\centering
\includegraphics[width=8cm]{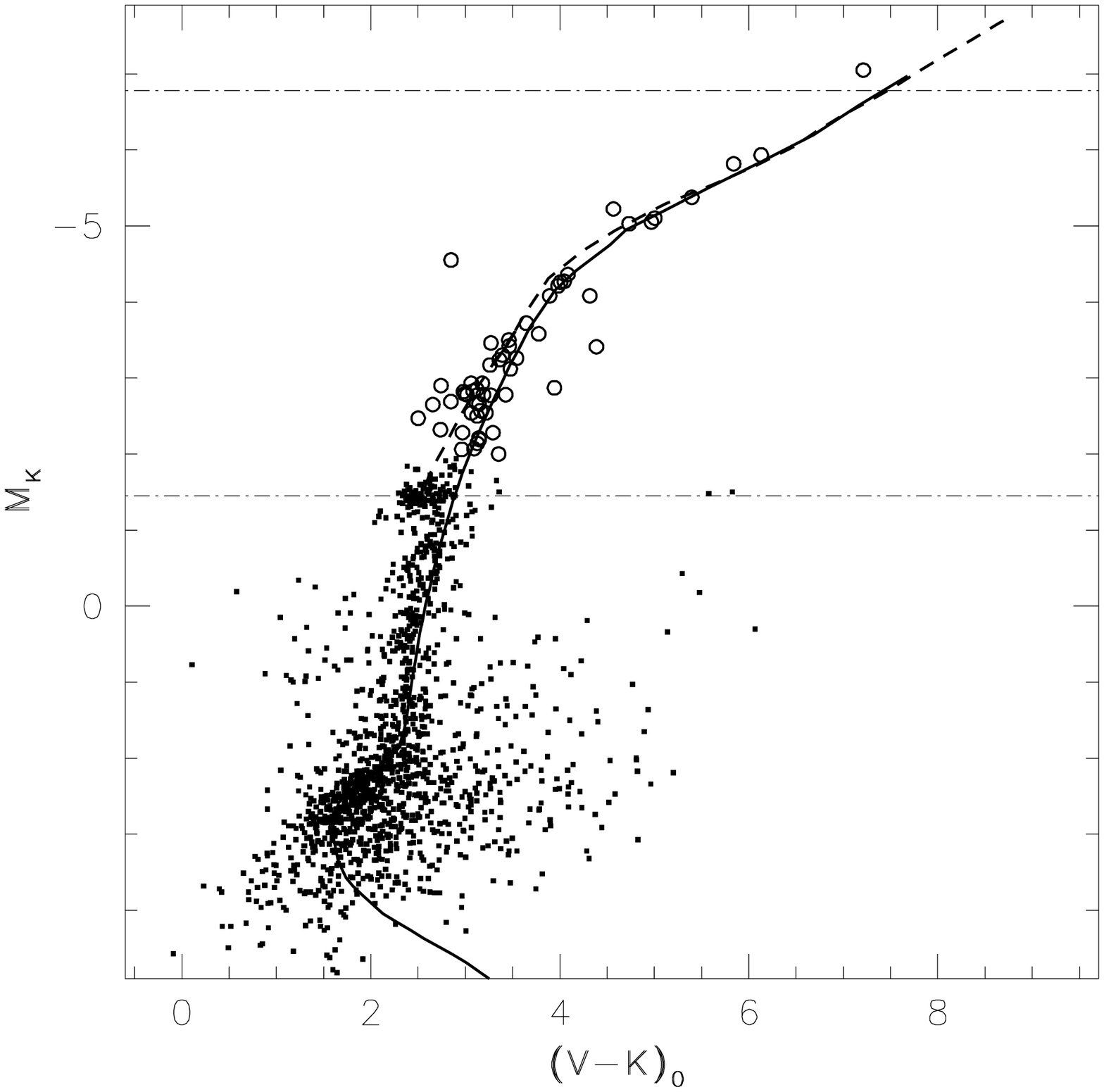}
\caption{The {\it cleaned} $M_K,(V-K)_{\circ}$ diagram of NGC\,6528 assuming
  $E_{\it   B-V}^{\rm  NGC6528}=0.55$   and   ($m-M$)$_{\circ}=14.44$.
  Superimposed   is  a theoretical  isochrone   from  Bertelli et  al.
  (\cite{bertelli94}) with  $12.6$  Gyr and  Z\,$= 0.02$.  A  dashed line
  highlights the HB and AGB phases.   The horizontal lines mark the HB
  and RGB  tip location.  Stars with $M_K<-2$  were drawn with different
  symbols for more visibility with respect to the isochrone.}
\label{f_vkk_bert}
\end{figure}


Figure~\ref{f_vkk_bert} shows      the     {\it      cleaned}   $M_K$,
($V-K$)$_{\circ}$  diagram of NGC\,6528 with  a $12.6$ Gyr and Z$=0.02$
theoretical isochrone from Bertelli et  al.  (\cite{bertelli94}).  The
best fit was obtained applying our values  for reddening and distance,
$E_{\it B-V}=0.55$ and ($m-M$)$_{\circ}=14.44$.  An even higher global
metallicity   of     Z$=$0.04    was    found   by    Feltzing      et
al. (\cite{feltzing01}) by using  the $\alpha$-enhanced  isochrones of
Salasnich  et al.    (\cite{salasnich00}).     Adopting  these  latter
isochrones to fit our $K$, $V-K$ diagrams would result in an extremely
high metallicity  of Z$=$0.07.  These isochrone  
would reconcile only 
with the   highest   estimates of metallicity   from
spectroscopy, such as those  by Carretta et  al.  (\cite{carretta01});
 we believe that  the  colours  of  these isochrones  may  be
systematically too blue.


%
%
%

\section{The $M_{\rm bol}$, $\log (T_{\rm  eff})$ diagram}

\begin{figure}[h]
\centering
\includegraphics[width=8cm]{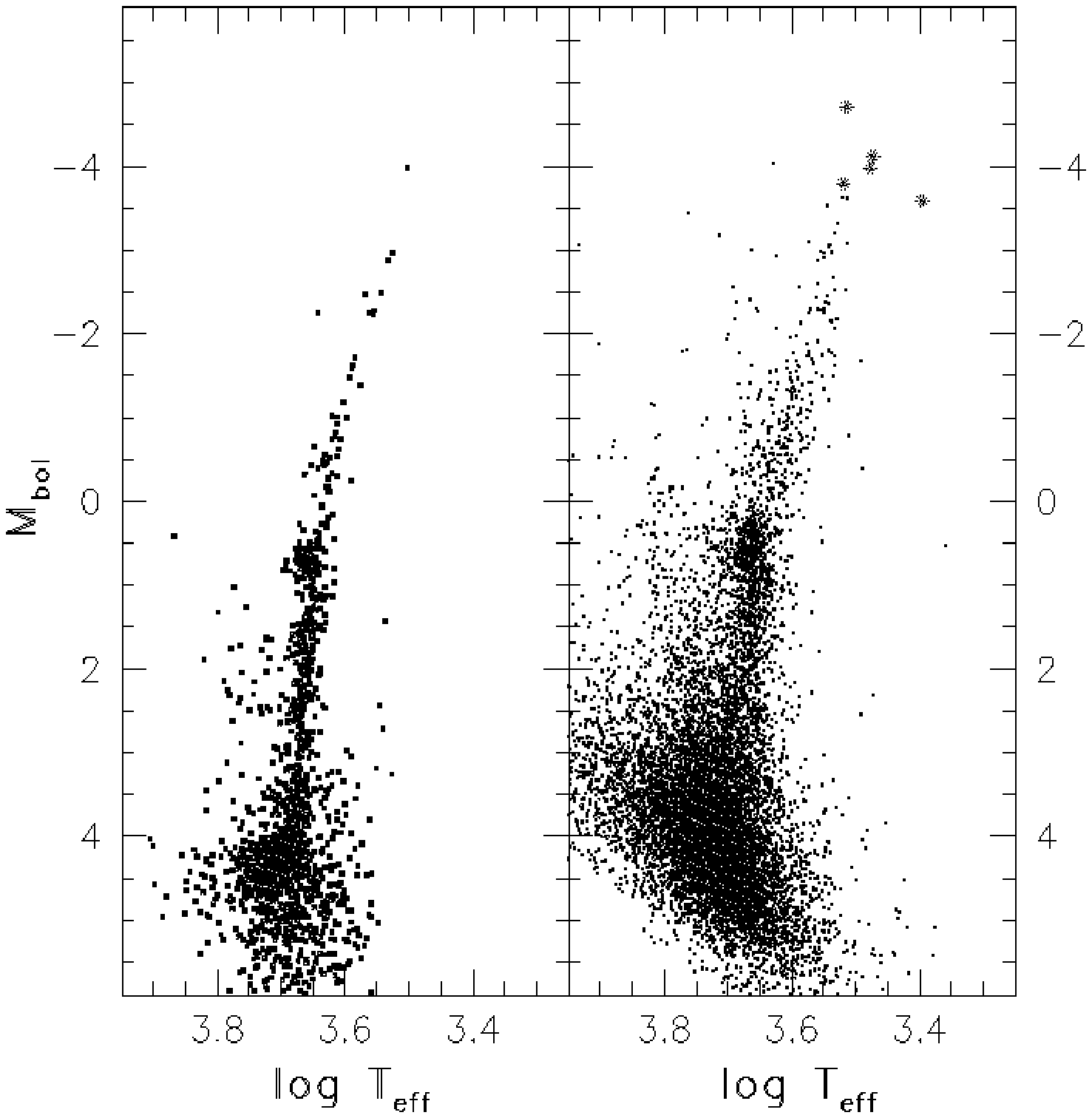}
\caption{The left panel  shows the {\it cleaned}  $K$, $V-K$ 
data    in the $M_{\rm bol}$,   log($T_{\rm  eff}$) theoretical plane.
Right panel shows the original  $K$,  $J-K$.  Starred symbols are  the
candidate AGB variables.}
\label{f_bolom_cmd}
\end{figure}

In Fig.~\ref{f_bolom_cmd} we  show  the NGC\,6528  data  in the 
$M_{\rm bol}$,  
$\log (T_{\rm eff})$  
plane.   Since  the decontamination process
strongly affects the brightest portion of the RGB (due to the combined
effects of saturation in the HST $V$ data and the smaller HST field of
view), we plot  both the cleaned  $K,V-K$ diagram (left panel) and the
original  $K, J-K$  data  (right panel).   Candidate AGB variables are
plotted as starred symbols.  The  observed CMD was converted into  the
$M_{\rm    bol}$, log($T_{\rm  eff}$)   plane   by  using  Table  4 of
Montegriffo  et al.   (\cite{montegriffo98}) to derive transformations
in  terms  of  the $V-K$ and $J-K$ colours.    
The bolometric corrections and effective temperature scales were
obtained by Montegriffo et al. as a function of NIR-optical colours,
using a large data base of giants in GCs with $-2.0<$[Fe/H]$<0.0$.
Using the cleaned diagram, we computed two of the F00 metallicity
indicators namely, Log $T_{\rm eff}^{\rm bump}$ and $M^{\rm bump}_{\rm
bol}$ (Table~\ref{t_ferraro}).


Although the membership of the variable  AGB candidates is still to be
confirmed spectroscopically, it  is  interesting to note   the analogy
with the bolometric magnitudes of the  confirmed long period variables
in  NGC\,6553,  where the  brightest     star  is found  at    $M_{\rm
bol}\simeq-4.7$.  Guarnieri et al.  (\cite{guarnieri97}) discussed the
maximum AGB luminosity as  an age indicator, and  used the NIR data of
NGC\,6553 (Guarnieri et al.  \cite{guarnieri98}) as a template for the
metal rich component  of external galaxies.  Interestingly, they found
that the brightest stars in NGC\,6553 were as  bright as the brightest
AGB stars  in the dwarf  elliptical  galaxy M32. Thus,  the metal rich
population of NGC\,6553, which is as old as  the Galactic halo GCs, is
able   to generate stars  as bright  as those  observed  in M32.  This
clearly shows that an ``intermediate age'' (few to several Gyr) is not
needed to account for the brightest stars in external galaxies such as
M32.

%
%

\section{Summary and conclusions}
\label{s_summary}

We have presented new high-quality   NIR observations of the  Galactic
globular cluster   NGC\,6528.  The $JHK_{\rm s}$  data allow
the construction of deep   NIR   CMDs reaching the   subgiant  branch.
Moreover, as in Feltzing et al.  (\cite{feltzing01}), we proper-motion
decontaminate  the  HST $V$ set  to  extract a clean  $VJHK$ sample.
Based on this wider colour baseline set, and on  a comparison with the
CMDs of NGC\,6553 and NGC\,104,  we derive new reddening and  distance
estimates    for      NGC\,6528:     $E_{\it   B-V}=0.55$          and
($m-M$)$_{\circ}=14.44$ (7.7 kpc).

We also test  various  calibrations  of photometric   metallicity
indicators.  Averaging  the results  from   10 metallicity 
indicators we
derive a mean value of  
[M/H] $\approx$ 0.0 for NGC\,6528.
The best  isochrone fit to our cleaned  $K$, $V-K$ diagram is obtained
adopting a $12.6$  Gyr and  Z $=0.02$ isochrone  from Bertelli  et al.
(\cite{bertelli94}).  
Thus, we conclude that the metallicity of NGC\,6528 is very
close to the mean of field stars of Baade's Window, [Fe/H]$=-0.25$, as
derived by MR94.

Although the membership of five AGB variable candidates is still to be
confirmed spectroscopically, these are bolometrically as bright as the
confirmed long period variables in NGC\,6553 and M32.  As discussed in
Guarnieri  et  al.    (\cite{guarnieri97}), this may    imply  that an
``intermediate age'' (few Gyr old)  component is not needed to account
for the brightest stars in external galaxies such as M32.

\begin{acknowledgements}
We thank  the anonymous referee for  helpful remarks that improved the
presentation of  this paper.  We also thank  Dr. Ferraro for providing
us their RGB fiducial  lines  and data, and  Dr.  Carretta  for useful
discussions.


\end{acknowledgements}

\end{document}